# 3D Imaging of a Phase Object from a Single Sample Orientation Using an Optical Laser

Chien-Chun Chen,[1] Huaidong Jiang,[2] Lu Rong,[1] Sara Salha,[1] Rui Xu,[1] Thomas G. Mason,[1,3] and Jianwei Miao[1*]

[1]*Department of Physics & Astronomy and California NanoSystems Institute, University of California, Los Angeles, CA 90095, USA.*

[2]*State Key Laboratory of Crystal Materials, Shandong University, Jinan 250100, People's Republic of China.*

[3]*Department of Chemistry and Biochemistry, University of California, Los Angeles, CA 90095, USA.*

**Ankylography is a new 3D imaging technique, which, under certain circumstances, enables reconstruction of a 3D object from a single sample orientation. Here, we provide a matrix rank analysis to explain the principle of ankylography. We then present an ankylography experiment on a microscale phase object using an optical laser. Coherent diffraction patterns are acquired from the phase object using a planar CCD detector and are projected onto a spherical shell. The 3D structure of the object is directly reconstructed from the spherical diffraction pattern. This work may potentially open the door to a new method for 3D imaging of phase objects in the visible light region. Finally, the extension of ankylography to more complicated and larger objects is suggested.**

## I. INTRODUCTION

Lens-based microscopies, such as light, phase-contrast, fluorescence, confocal, x-ray and electron, have made important contributions to a broad range of fields in both physical and life sciences. In 1999, a new form of microscopy was developed, termed coherent diffraction imaging or coherent diffraction microscopy [1], in which the diffraction pattern of a non-crystalline specimen was first measured and then directly phased to obtain an image. The well-known phase problem was solved by oversampling the diffraction intensity [2,3] in combination of iterative algorithms [4-7]. Using synchrotron radiation, high harmonic generation, soft x-ray laser sources and free electron lasers, coherent diffraction imaging has been applied to conduct structure studies of a wide range of samples in materials science, nanoscience and biology [8-31]. To perform 3D coherent diffraction imaging (CDI), it is necessary to acquire a sequence of 2D diffraction patterns by either tilting a sample at multiple orientations or using many identical copies of the sample [9,12,15,25,30]. In some applications, however, it is very desirable to obtain the 3D structure of an object without requiring sample tilting or multiple copies. To achieve this challenging goal, ankylography has recently been developed [32], which under certain circumstances allows for 3D imaging of an object from a single sample orientation. Subsequently, two imaging methods that are somewhat related to ankylography have been demonstrated. The first is super-resolution biomolecular crystallography [33], which under some conditions can determine the high-resolution 3D structure of macromolecules from low-resolution data. The other is discrete tomography [34], which enables to achieve the 3D atomic reconstruction of a small crystalline nanoparticle by only using two projections, combined with prior knowledge of the particle's lattice structure. Compared to conventional 3D structure and imaging methodology, these three methods are mathematically ill-posed problems, but represent a new and important direction in structural determination – retrieving 3D structural information from a portion of Fourier magnitudes or coefficients.

In this paper, we first provide a matrix rank analysis to explain why ankylography under certain circumstances can be used to determine the 3D structure from a single sample orientation. We then perform the ankylographic reconstruction of a phase object using an optical laser. There are three significant implications of this experiment. First, it extends ankylography to the 3D imaging of phase objects in the visible light region that is currently dominated by confocal microscopy. Second, compared to the previous result that is somewhat controversial due to the use of a transparent sample on an opaque substrate [35,36], this work represents the first ankylographic reconstruction of a phase object on a transparent substrate. Finally, using X-ray free electron lasers, ankylography may be applied to determine the 3D structure of certain classes of samples without the need of identical copies.

## II. MATRIX RANK ANALYSIS OF ANKYLOGRAPHY

We provide a matrix rank analysis to explain why ankylography under certain circumstances can be used to determine the 3D structure from a single view. Let us assume that a coherent wave illuminates a 3D real object, $\rho(x,y,z)$. The far-field diffracted wave, $F(k_x,k_y,k_z)$, is oversampled on a spherical shell. We separate $F(k_x,k_y,k_z)$ into cosines and sines,

$$
\begin{aligned}
&F(k_x,k_y,k_z) = A_{k_x,k_y,k_z}\exp(i\phi_{k_x,k_y,k_z}) \\
&= \sum_{x=-M}^{M}\sum_{y=-M}^{M}\sum_{z=-M}^{M}\rho(x,y,z)\exp\left[\frac{-2\pi i(k_x\cdot x+k_y\cdot y+k_z\cdot z)}{2N+1}\right] \\
&\Rightarrow\begin{cases}
A_{k_x,k_y,k_z}\cos(\phi_{k_x,k_y,k_z}) = \displaystyle\sum_{x=-M}^{M}\sum_{y=-M}^{M}\sum_{z=-M}^{M}\rho(x,y,z)\cos\left[\frac{2\pi(k_x\cdot x+k_y\cdot y+k_z\cdot z)}{2N+1}\right] \\
iA_{k_x,k_y,k_z}\sin(\phi_{k_x,k_y,k_z}) = -i\displaystyle\sum_{x=-M}^{M}\sum_{y=-M}^{M}\sum_{z=-M}^{M}\rho(x,y,z)\sin\left[\frac{2\pi(k_x\cdot x+k_y\cdot y+k_z\cdot z)}{2N+1}\right]
\end{cases} \\
&\qquad\forall\, k_x,k_y,k_z:\ \left(N-\frac{1}{2}\right)^2 \le k_x^2+k_y^2+(k_z+N)^2 < \left(N+\frac{1}{2}\right)^2
\end{aligned}
\tag{1}
$$

where $(2M+1)^3$ is the size of the 3D object (*i.e.* support size), $(2N+1)^3$ is the size of the Fourier-space array in which the two hemi-spherical shells are located, $A_{k_x,k_y,k_z}$ and $\phi_{k_x,k_y,k_z}$ are the magnitudes and phases of $F(k_x,k_y,k_z)$, and the diffraction angle is assumed to be 90°. In Eq. (1), we chose the spherical shell to be one voxel thick, which is a reasonable assumption as the thickness of the spherical shell is determined by the experimental parameters such as the energy resolution, divergence and convergence angle of the incident beam. Note that Eq. (1) is not the discrete Fourier transform relation as the reciprocal-space vectors on the spherical shell ($k_x$, $k_y$, $k_z$) are not independent, but related via $(N-1/2)^2 \le k_x^2+k_y^2+(k_z+N)^2 < (N+1/2)^2$. We rewrite Eq. (1) into a matrix form,

$$BX = A$$

$$B = \begin{pmatrix} \cos\left[\frac{2\pi(k_{x_1}\cdot x_1 + k_{y_1}\cdot y_1 + k_{z_1}\cdot z_1)}{2N+1}\right] & \cdots & \cos\left[\frac{2\pi(k_{x_1}\cdot x_{(2M+1)^3} + k_{y_1}\cdot y_{(2M+1)^3} + k_{z_1}\cdot z_{(2M+1)^3})}{2N+1}\right] \\ \vdots & \vdots & \vdots \\ \cos\left[\frac{2\pi(k_{x_L}\cdot x_1 + k_{y_L}\cdot y_1 + k_{z_L}\cdot z_1)}{2N+1}\right] & \cdots & \cos\left[\frac{2\pi(k_{x_L}\cdot x_{(2M+1)^3} + k_{y_L}\cdot y_{(2M+1)^3} + k_{z_L}\cdot z_{(2M+1)^3})}{2N+1}\right] \\ 1 & \cdots \quad \cdots & 1 \\ -\sin\left[\frac{2\pi(k_{x_1}\cdot x_1 + k_{y_1}\cdot y_1 + k_{z_1}\cdot z_1)}{2N+1}\right] & \cdots & -\sin\left[\frac{2\pi(k_{x_1}\cdot x_{(2M+1)^3} + k_{y_1}\cdot y_{(2M+1)^3} + k_{z_1}\cdot z_{(2M+1)^3})}{2N+1}\right] \\ \vdots & \vdots & \vdots \\ -\sin\left[\frac{2\pi(k_{x_L}\cdot x_1 + k_{y_L}\cdot y_1 + k_{z_L}\cdot z_1)}{2N+1}\right] & \cdots & -\sin\left[\frac{2\pi(k_{x_L}\cdot x_{(2M+1)^3} + k_{y_L}\cdot y_{(2M+1)^3} + k_{z_L}\cdot z_{(2M+1)^3})}{2N+1}\right] \end{pmatrix} \quad (2)$$

$$X = \begin{pmatrix} \rho(x_1, y_1, z_1) \\ \vdots \\ \rho(x_{(2M+1)^3}, y_{(2M+1)^3}, z_{(2M+1)^3}) \end{pmatrix} \qquad A = \begin{pmatrix} A_1\cos(\phi_1) \\ \vdots \\ A_L\cos(\phi_L) \\ A_0 \\ A_1\sin(\phi_1) \\ \vdots \\ A_L\sin(\phi_L) \end{pmatrix}$$

$$\forall\, k_{x_i}, k_{y_i}, k_{z_i} \quad x_i, y_i, k_i \in [-N, N]:\ \left(N - \frac{1}{2}\right)^2 \le k_{x_i}^2 + k_{y_i}^2 + (k_{z_i} + N)^2 < \left(N + \frac{1}{2}\right)^2$$

where *B*, *X* and *A* are $(2L+1)\times(2M+1)^3$, $(2M+1)^3\times 1$ and $(2L+1)\times 1$ matrices, respectively, $(L+1)$ is the number of non-centro-symmetrical grid points on the spherical shell, and the row of (1 … 1) in matrix *B* and $A_0$ in matrix *A* correspond to the centro-voxel. To facilitate our quantitative analysis, we generate two new matrices $B'$ and $X'$ by expanding *B* and padding zeros to *X*,

$$B' = \begin{pmatrix} \cos\left[\frac{2\pi(k_{x_1}\cdot x_1 + k_{y_1}\cdot y_1 + k_{z_1}\cdot z_1)}{2N+1}\right] & \cdots & & \cdots\cos\left[\frac{2\pi(k_{x_1}\cdot x_{2L+1} + k_{y_1}\cdot y_{2L+1} + k_{z_1}\cdot z_{2L+1})}{2N+1}\right] \\ \vdots & & B & \vdots \\ -\sin\left[\frac{2\pi(k_{x_L}\cdot x_1 + k_{y_L}\cdot y_1 + k_{z_L}\cdot z_1)}{2N+1}\right] & \cdots & & \cdots -\sin\left[\frac{2\pi(k_{x_L}\cdot x_{2L+1} + k_{y_L}\cdot y_{2L+1} + k_{z_L}\cdot z_{2L+1})}{2N+1}\right] \end{pmatrix}$$

$$X' = \begin{pmatrix} 0 \\ \vdots \\ 0 \\ X \\ 0 \\ \vdots \\ 0 \end{pmatrix} \qquad \textit{such that} \qquad B'X' = A \tag{3}$$

where $B'$ is defined as the sampling matrix, $B'$ and $X'$ are $(2L+1)\times(2L+1)$ and $(2L+1)\times 1$ matrices, respectively. Mathematically, Eq. (3) is equivalent to Eq. (2).

To give some specific examples on the matrix rank analysis, we first calculated the rank of $B'$ by using a 7×7×7 array (*i.e.* $M$ = 3). The spherical shell is embedded inside a 17×17×17 array (*i.e.* $N$ = 8). The number of non-centro-symmetrical grid points on the spherical shell of 1 voxel thick is 393 (*i.e.* $L$ = 392) with the oversampling degree ($O_d$ = 1.14), defined as [32]:

$$O_d = \frac{\textit{Number of voxels within one of the spherical shell}}{\textit{Number of voxels within the support}}. \tag{4}$$

The rank of $B'$ is determined to be 785 (*i.e.* matrix $B'$ has full rank) with tolerance of $10^{-3}$. In this case, the number of unknown variables of the 3D object is 343 (*i.e.* $7^3$), and the number of unknown variables for the phases in Eq. (3) is 392. Therefore the total number of unknown variables is smaller than the rank of $B'$, suggesting that the 3D object can in principle be obtained by solving Eq. (3). We also calculate the rank of $B'$ for a 14×14×14 voxel object with $O_d$ = 2.06. In this case, the rank of $B'$ is larger than the number of unknown variables with tolerance of $10^{-6}$, but smaller with tolerance of $10^{-3}$. When $O_d$ is increased to be ~4.0, the rank of B′ (with tolerance of $10^{-3}$) is larger than the number of unknown variables. The matrix rank analysis suggests that when the object array is larger, the tolerance becomes smaller in order to maintain full rank of the sampling matrix, and the ankylographic reconstruction becomes more challenging without additional constraints and information, which is consistent with the numerical simulation results [32]. To facilitate interested readers who might wish to conduct ankylographic reconstructions, several Matlab source codes have been posted on a public website and can be freely downloaded to test this method [37].

## III. ANKYLOGRAPHY EXPERIMENT AND RECONSTRUCTION

Next, we present an ankylographic experiment on a phase object using an optical laser. Figure 1 shows the schematic layout of the experimental setup. An optical laser with $\lambda$ = 543 nm was collimated by a compound lens system, consisting of two converging lenses and producing a parallel beam with a diameter of ~200 μm. An aperture was placed 15 mm upstream of the sample to block the unwanted scattering from the lenses. The object to be imaged in 3D is a dielectric phase pattern made up of non-absorbing SU-8 epoxy photoresist that had been cross linked by using an Ultratech XLS stepper. Figure 2(a) shows a differential-interference-contrast (DIC) microscope image of the phase object, which consists of a dense raft-like arrangement of four alphabet letters (WWWA) in close proximity; as fabricated, each plate-like letter is about 4 μm wide x 7 μm tall x 1 μm thick with ≈ 1 μm effective pen width [38,39]. As the sample is a weak phase object, the phase shift within a 3D resolution volume can be approximately represented as

$$e^{i\varphi(x,y,z)} = 1 + i\varphi(x,y,z)\,. \qquad (5)$$

The Fourier transform of the term "1" in Eq. (5) is concentrated at the center voxel in reciprocal space (*i.e.* the direct wave) and is blocked by a beamstop, while the Fourier modulus of $i\varphi(x,y,z)$ is centro-symmetrical. Compared to a conventional 2D exit wave, where the phase shift $\varphi(x,y)$ may not be small after propagating through a whole object, $\phi(x,y,z)$ represents the phase shift within 1 voxel in ankylography and is thus small for a weak phase object. The sample was supported on a silicon nitride membrane of 100 nm thickness. To increase the depth of the sample along the Z (beam) axis, the silicon nitride membrane was tilted about 45° relative to the incident beam. Coherent diffraction patterns were recorded by a liquid-nitrogen-cooled CCD camera with 1340×1300 pixels and a pixel size of 20 μm × 20 μm, positioned at a distance of 31.5 mm from the sample. The distance between the sample and the detector could not be further reduced due to the geometry of the CCD camera. A beamstop was positioned in front of the CCD camera to block the direct beam.

To obtain coherent diffraction patterns at highest possible resolution, we moved the CCD camera both horizontally and vertically, and measured a diffraction pattern at each of the four quadrants. The four diffraction patterns were tiled together to form a high spatial resolution (HSR) pattern. To ensure the missing center confined within the centro-speckle [40], we took an additional low spatial resolution (LSR) diffraction pattern by moving the CCD camera further downstream at a distance of 108 mm to the sample. To remove the background scattering and readout noise of the CCD, we measured two sets of diffraction patterns at each position with the sample in and out of the laser beam. Table 1 shows the experimental parameters used to measure the diffraction patterns. The HSR and LSR diffraction patterns after background subtraction are shown in Figs. 2(b) and (c), which were combined to assemble a diffraction pattern of 2001×2001 pixels with a small missing center.

Because the CCD is a 2D planar detector, the assembled diffraction pattern has to be projected onto a spherical surface. As the solid angle subtended by each CCD pixel varies with the diffraction angle, the diffraction intensity was normalized by

$$I_N\left(k_x^d,k_y^d\right)=\frac{\Delta\Omega(0,0)}{\Delta\Omega\left(k_x^d,k_y^d\right)}I_M\left(k_x^d,k_y^d\right) \qquad (6)$$

where $I_N(k_x^d,k_y^d)$ and $I_M(k_x^d,k_y^d)$ are the normalized and measured diffraction intensities, $(k_x^d,k_y^d)$ is the pixel position of the planar CCD, $\Delta\Omega(0,0)$ and $\Delta\Omega\left(k_x^d,k_y^d\right)$ are the solid angle subtended by the central pixel and pixel $(k_x^d,k_y^d)$, respectively. $\Delta\Omega(k_x^d,k_y^d)$ is determined by,

$$\Delta\Omega\left(k_x^d,k_y^d\right)=R\int_{k_x^d-\delta/2}^{k_x^d+\delta/2}\int_{k_y^d-\delta/2}^{k_y^d+\delta/2}\frac{dk_x dk_y}{[(k_x)^2+(k_y)^2+R^2]^{3/2}} \qquad (7)$$

where $R$ is the distance from the sample to the CCD camera and $\delta$ is the CCD pixel size.

The normalized diffraction intensity was then projected onto the spherical surface on a Cartesian grid. To perform more accurate interpolation, we first located the Cartesian grid points, ($k_x^c,k_y^c,k_z^c$), within a spherical shell of 1 voxel thick and then projected the grid points onto the planar CCD by

$$k_x^d{}'=R\frac{k_x^c}{R-k_z^c} \qquad k_y^d{}'=R\frac{k_y^c}{R-k_z^c}, \qquad (8)$$

where $(k_x^d{}', k_y^d{}')$ are the X and Y coordinates on the detector plane and are not necessarily an integer number of pixels. We calculated $I_N(k_x^d{}', k_y^d{}')$ using spline interpolation with the neighboring pixels, and then assigned $I_N(k_x^d{}', k_y^d{}')$ to the Cartesian grid point, $I_N(k_x^c, k_y^c, k_z^c)$. Figure 2(d) shows the diffraction intensity distributed within two spherical shells on a 3D Cartesian grid. The centro-symmetry of the diffraction intensity is because the sample is a weak phase object (Eq. (5)). The array size of the 3D Cartesian grid is 1691×1691×491 voxels with a diffraction angle of 32.3°.

To perform the ankylographic reconstruction, we first roughly estimated a loose support for the phase object. The algorithm was then iterated back and forth between real and reciprocal space with a random phase set as an initial input. In real space, the voxel value outside the support and the negative voxel value inside the support were slowly pushed close to zero [6]. In reciprocal space, the Fourier magnitudes within the spherical shell were updated with the measure ones while other Fourier magnitudes remained unchanged in each iteration. The convergence of the algorithm was monitored by an $R_{sphere}$ defined as,

$$R_{sphere} = \frac{\left\| F^M_{sphere}(\vec{k}) | - | F^C_{sphere}(\vec{k}) \right\|}{| F^M_{sphere}(\vec{k}) |} \qquad (9)$$

where $| F^M_{sphere}(\vec{k}) |$ and $| F^C_{sphere}(\vec{k}) |$ are the measured and calculated Fourier modulus within a spherical shell. Compared to phase retrieval in coherent diffraction imaging, the convergence speed in ankylographic reconstruction is slower and more iterations are required. To make ankylographic reconstructions more efficient, we performed ~10 independent reconstructions each with a random phase seed. After 5000 iteration, we chose the best 3D reconstruction with the smallest $R_{sphere}$. By convolving the reconstruction with a Gaussian filter and choosing a cutoff value, we determined an updated support. After running another 5000 iterations, we reconstructed a 3D object from which a final tight support was determined. Figure 3 shows the supports used from loose to tight during ankylographic reconstructions. The oversampling degree ($O_d$) for the final support is 2062 [32]. Such a large oversampling degree in the reconstruction occurs because the final support that we used is very tight. After another 5000 iterations, a final 3D reconstruction was obtained, corresponding to $R_{sphere}$ = 0.36. According to our experience, enforcing a correct, tight support is important in ankylographic reconstruction. In addition, a larger oversampling degree ($O_d$) also improves reconstruction of experimental data.

## IV. RESULTS

The resolution in ankylography is determined by $d_t = \lambda / \sin(2\theta)$ and $d_l = \lambda / (2\sin^2\theta)$, where $d_t$ and $d_l$ represent the transverse and longitudinal resolution (*i.e.* perpendicular and parallel to the incident beam), $\lambda$ is the wavelength and $2\theta$ is the diffraction angle. In this experiment, the transverse and longitudinal resolution was estimated to be ~1.0 μm and ~3.5 μm, respectively. Figures 4(a-f) show 3 projections and 3 central slices of the final reconstruction along the X, Y, and Z (beam) axes. Based on the achieved resolution of ~1.0 μm along the X and Y axes and ~3.5 μm along the Z axis, we determined the projection length of the object in the X, Y and Z axes to be ~19 μm, ~23 μm and ~23 μm, respectively. Figure 5(a) shows an iso-surface rendering of the ankylographic reconstruction, and the orientation of the phase object relative to the incident beam. To verify the reconstruction, we tilted the reconstruction to the same orientation (Fig. 5b) as shown in the differential interference contrast (DIC) image (Fig. 2a). The 3 letters "WWW" are clearly visible and consistent with the DIC image, while the letter "A" is a bit too small to be resolved in the reconstruction. To further quantify the ankylographic reconstruction, we performed a line scan across the reconstruction (Fig. 5b). The blue curve in Fig. 5(c) shows the reconstructed density of the phase object, which is in reasonably good agreement with the DIC curve (in red). Differences in the appearance of the two images are expected because ankylography, when applied in an optical context to a structured, non-absorbing, dielectric material, produces a quantitative reconstruction of the density of dielectric polarizability of a phase object, not a DIC image that incorporates interference effects.

## V. CONCLUSION

In this article, we have presented a matrix rank analysis to explain why ankylography, under certain circumstances, enables reconstruction of a 3D object from a single spherical diffraction pattern. We have demonstrated this approach by performing an ankylography experiment on a dielectric phase object using an optical laser. Coherent diffraction patterns were measured from the phase object, projected onto a spherical surface, and directly phased to obtain the 3D structure of the object. Transverse and longitudinal resolutions of 1.0 μm and 3.5 μm, respectively, were achieved in the experiment. While the resolution is currently limited by the experimental set-up (*i.e.* the distance between the sample and the CCD could not be set smaller than 31.5 mm due to the geometry of the CCD camera), the ultimate resolution is set by the wavelength of the incident beam. Thus, we anticipated that even better resolution can be achieved in future experiments.

Compared to conventional coherent diffraction imaging [8-31], the ankylographic reconstruction not only requires a tight support with a large oversampling degree, but also becomes more challenging for larger objects. In order to apply ankylography to large objects, three different approaches are envisioned. First, our numerical simulations suggest that increasing the thickness of the spherical shell can distinctly improve the ankylographic reconstruction of large objects. Experimentally, this may be realized by using an incident wave with an energy bandwidth, coupled with an energy-resolved detector [41]. Second, more real-space constraints can facilitate the ankylographic reconstruction of large objects. One way to achieve this is to position a 3D object with a known structure close to an unknown one, which is somewhat related to molecular replacement and holography [42,43]. Based on our numerical simulations, the combination of the known part and a spherical diffraction pattern is more effective in reconstructing a large 3D object. Finally, by acquiring several spherical diffraction patterns at different sample orientations with each having a large oversampling degree, our numerical simulations indicate that ankylography can be extended to larger objects. Compared to conventional tomography, the number of projections required in ankylography will likely be smaller due to the utilization of spherical diffraction patterns.

This work was in part supported by the U.S. Department of Energy, Office of Basic Energy Sciences (DE-FG02-06ER46276) and the U.S. National Institute of Health (GM081409-01A1). H. J. is supported by NSFC (51002089), and Independent Innovation Foundation of Shandong University (2010JQ004).

*Corresponding author (Email: miao@physics.ucla.edu).

| | | Sample (exposure time × number of frames) | Background (exposure time × number of frames) | Distance from sample to CCD |
|---|---|---|---|---|
| HSR | Center | 0.17 s × 1000 | 0.17 s × 500 | 3.15 cm |
| | Lower-Left | 0.18 s × 1000 | 0.18 s × 500 | |
| | Lower-Right | 0.45 s × 1000 | 0.45 s × 500 | |
| | Upper-Left | 0.2 s × 1000 | 0.2 s × 500 | |
| | Upper-Right | 0.16 s × 1000 | 0.16 s × 500 | |
| LSR | | 0.25 s ×1000 | 0.25 s × 500 | 10.80 cm |

**Tab. 1** Experimental parameters used to measure the high spatial resolution (HSR) and low spatial resolution (LSR) diffraction patterns using an optical laser ($\lambda$ = 543 nm). The incident flux on the sample was estimated to be ~$1.7\times10^8$ photons/$\mu m^2 \cdot s$.

**Figure Captions**

**Figure 1** Schematic layout of the experimental set-up. A compound lens system, consisting of two converging lenses, was used to collimate the incident laser beam with a wavelength of 543 nm. An aperture was placed 15 mm upstream of the sample to block the unwanted scattering from the lenses. A phase object made up of SU-8 epoxy photoresist was supported on a silicon nitride membrane of 100 nm thick. To increase the depth of the sample along the beam axis, the silicon nitride membrane was tilted about 45° relative to the incident beam. Coherent diffraction patterns were recorded by a liquid-nitrogen-cooled CCD camera with 1340×1300 pixels and a pixel size of 20 μm×20 μm, placed at a distance of 31.5 mm from the sample. A beamstop was positioned in front of the CCD camera to block the direct beam.

**Figure 2** (**a**) DIC microscope image of the phase object, consisting of four alphabet letters (WWWA). (**b**), (**c**) The high and low spatial resolution diffraction patterns acquired by a planar CCD detector. The low spatial resolution pattern was used to reduce the missing center. (**d**) Two spherical diffraction patterns on a 3D Cartesian grid. The centro-symmetry of the two spherical patterns is because the sample is a phase object. The size of the 3D array is 1691×1691×491 voxels with a diffraction angle of 32.3°.

**Figure 3** Supports from loose (**a**) to tight (**c**) used for the ankylographic reconstructions. (**a**) Initial loose support. (**b**) Updated support, (**c**) Final tight support.

**Figure 4** . (**a-c**) Three projections of the final reconstruction along the X, Y, and Z (beam) axes. Based on the achieved resolution of ~1.0 μm along the X and Y axes and ~3.5 μm along the Z axis, the projection length of the object in the X, Y and Z axes was estimated to be ~19 μm, ~23 μm and ~23 μm, respectively. (**d-f**) Three central slices of the final reconstruction along the X, Y and Z axes.

**Figure 5** (**a**) Iso-surface rendering of the ankylographic reconstruction of the phase object where the relative orientation of the incident beam to the object position is illustrated. (**b**) The reconstruction is tilted to the same orientation as the DIC image (Fig. 2a). Although the resolution of the reconstruction is lower than the DIC image, the two images are in good agreement. (**c**) Line scans across the reconstruction and the DIC image. The two curves agree reasonably well. The discrepancy is ankylography produces a quantitative reconstruction of the phase object, but not the DIC image.

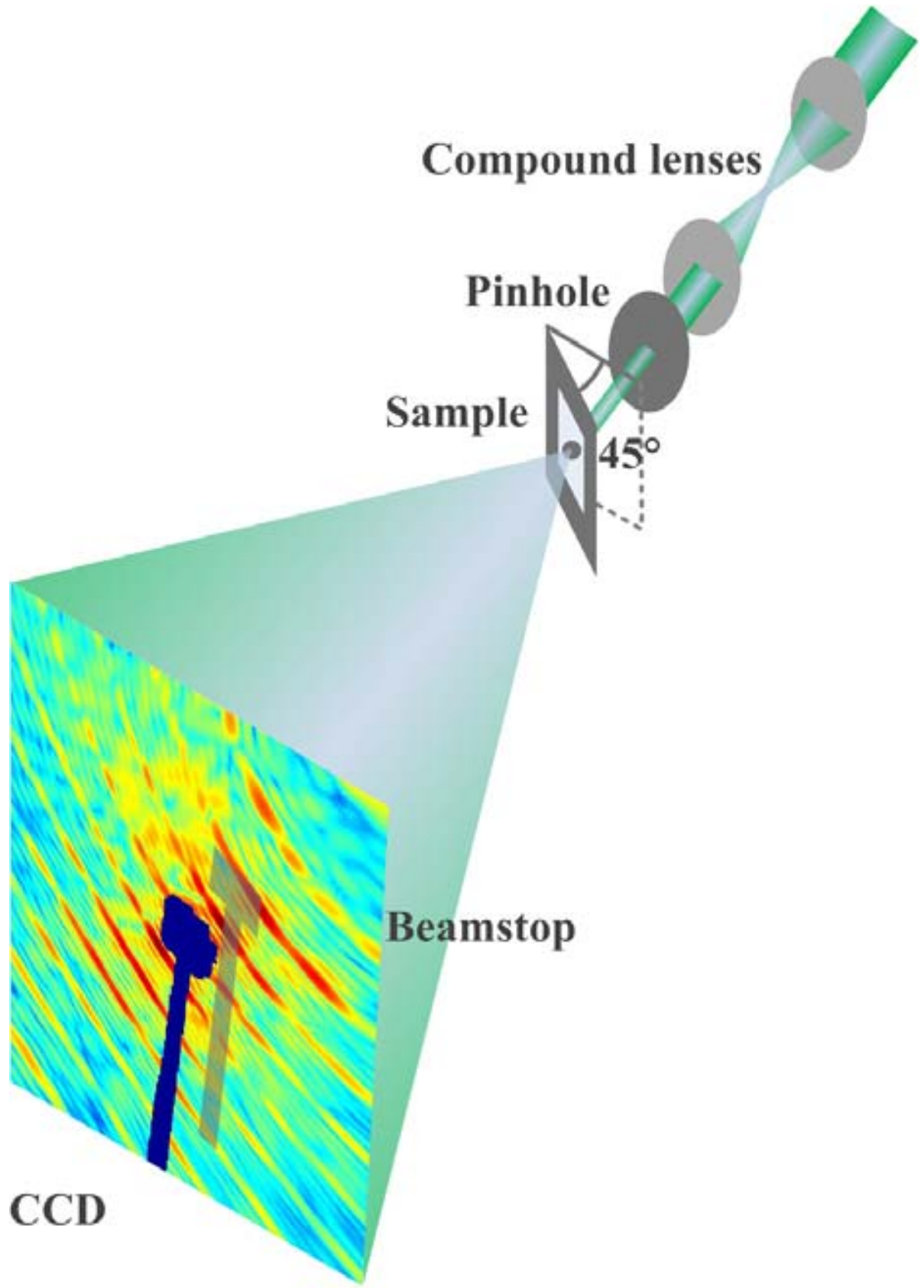

**FIG. 1**

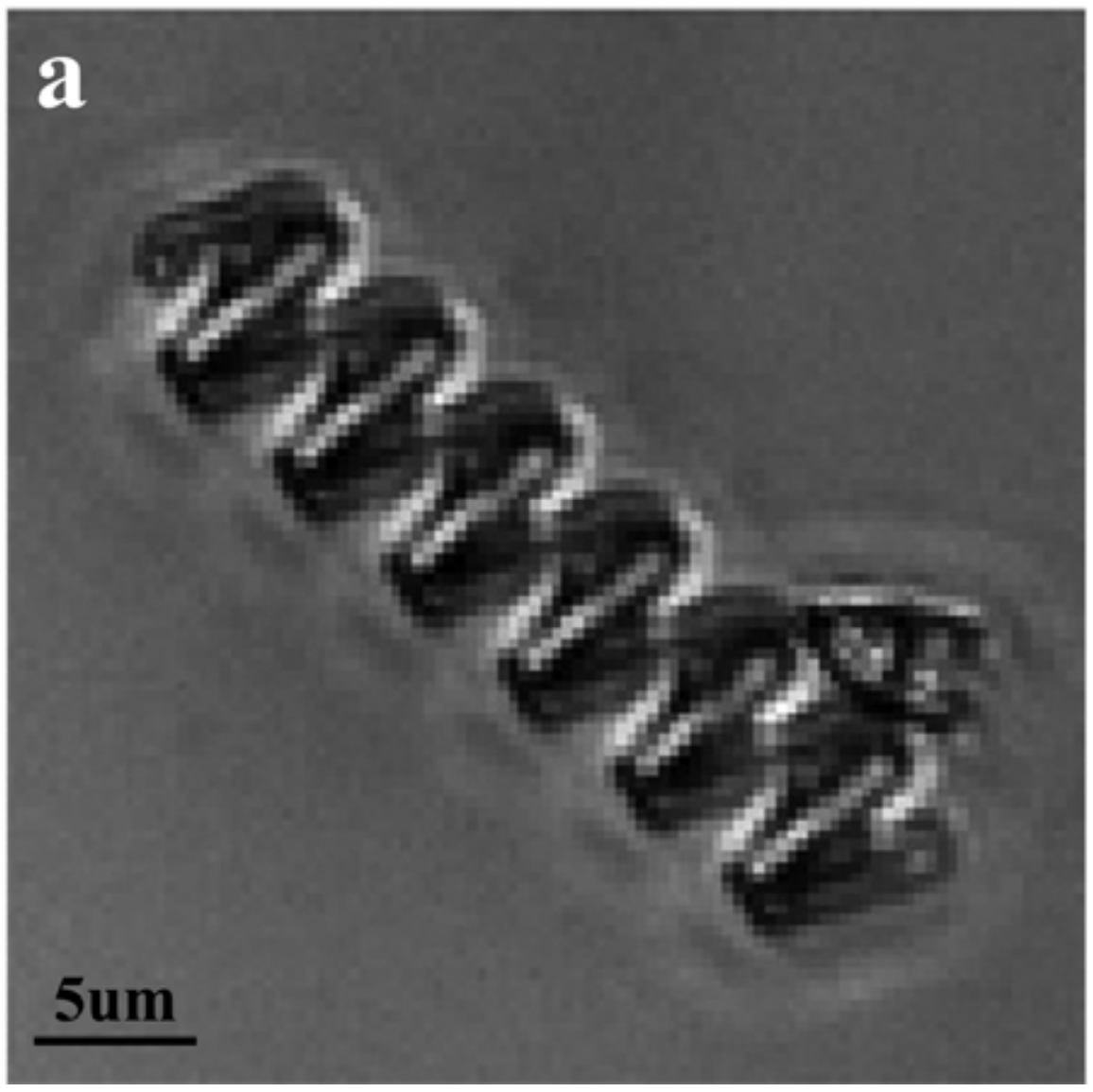

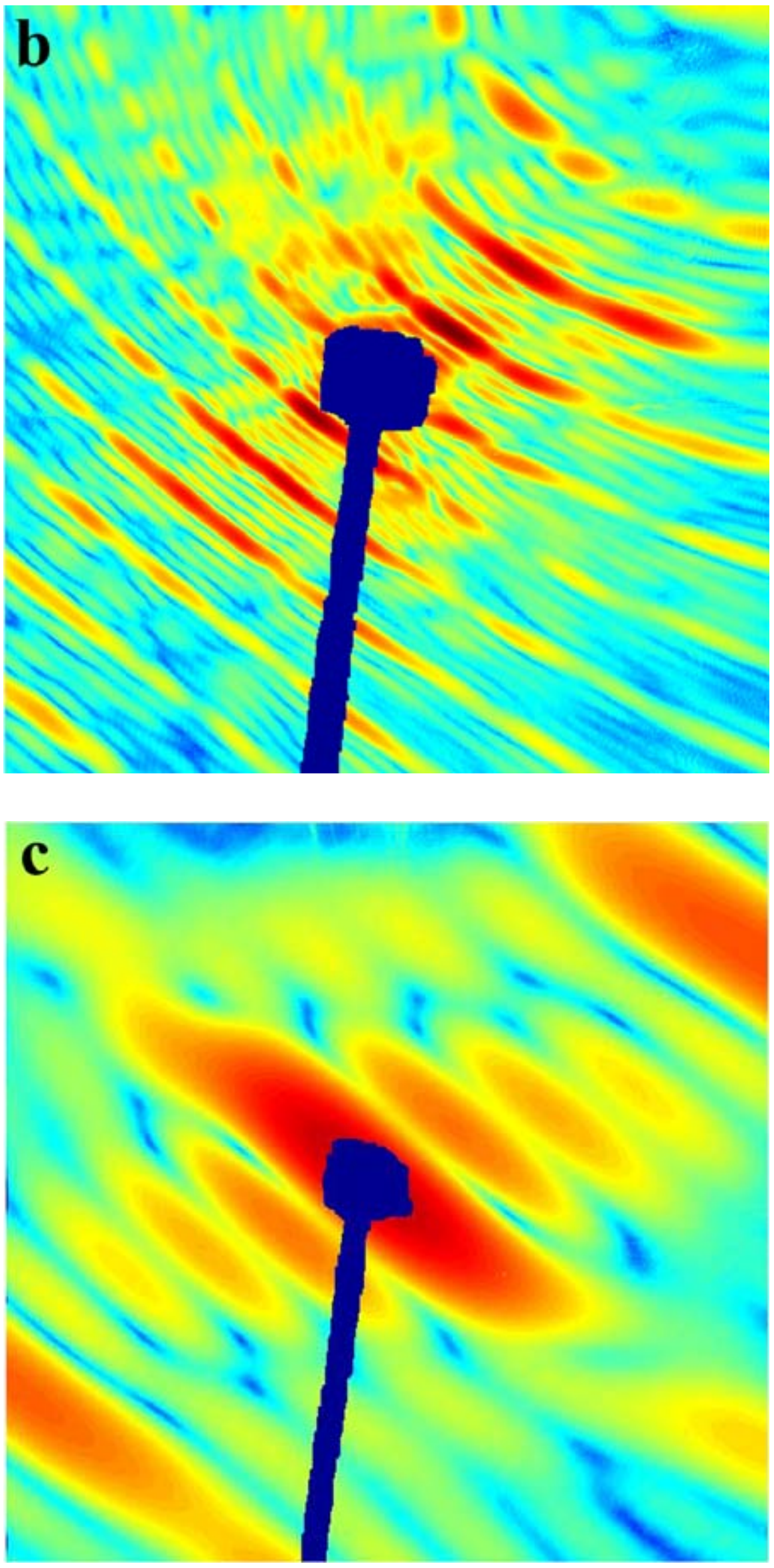
b
c

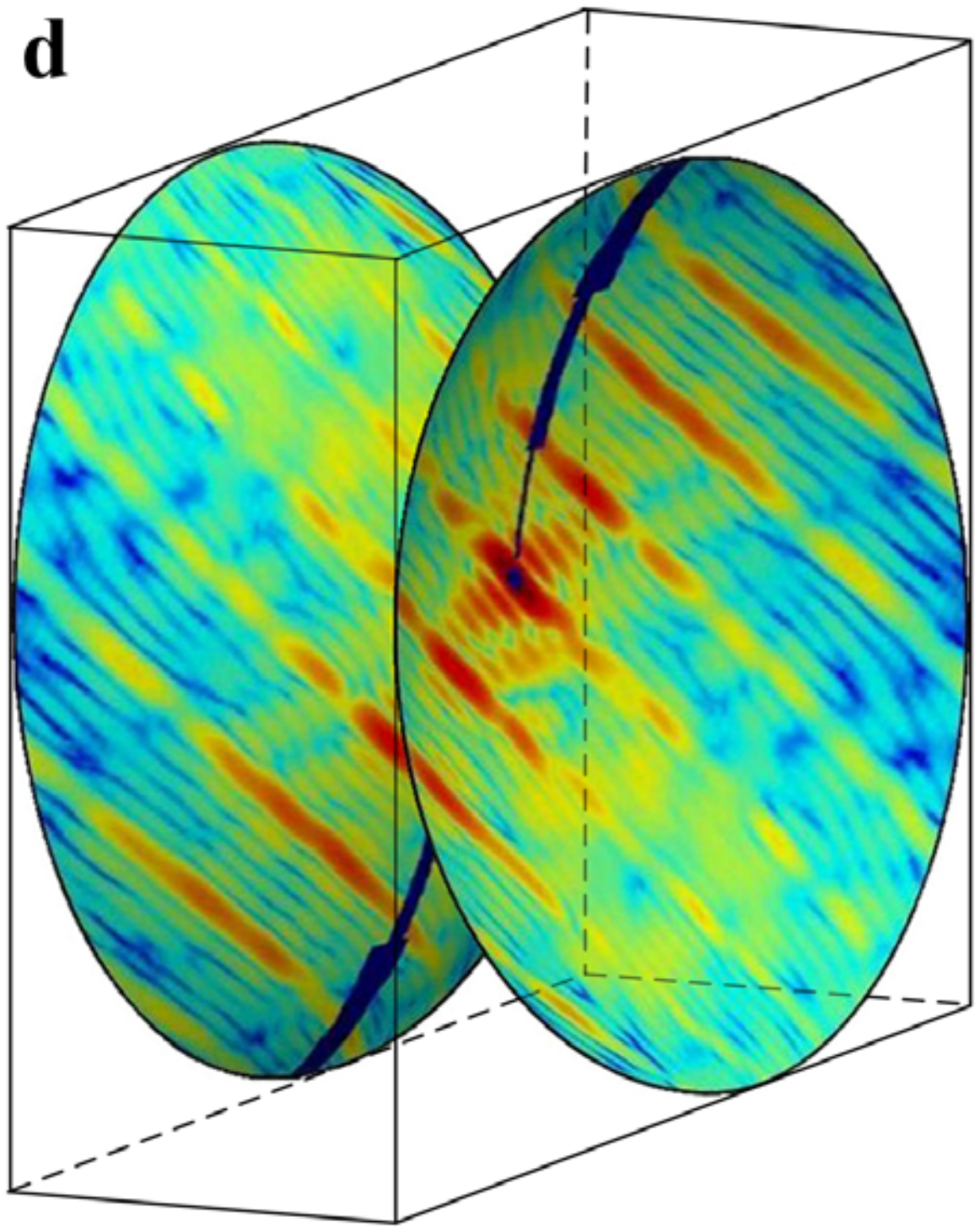

FIG. 2

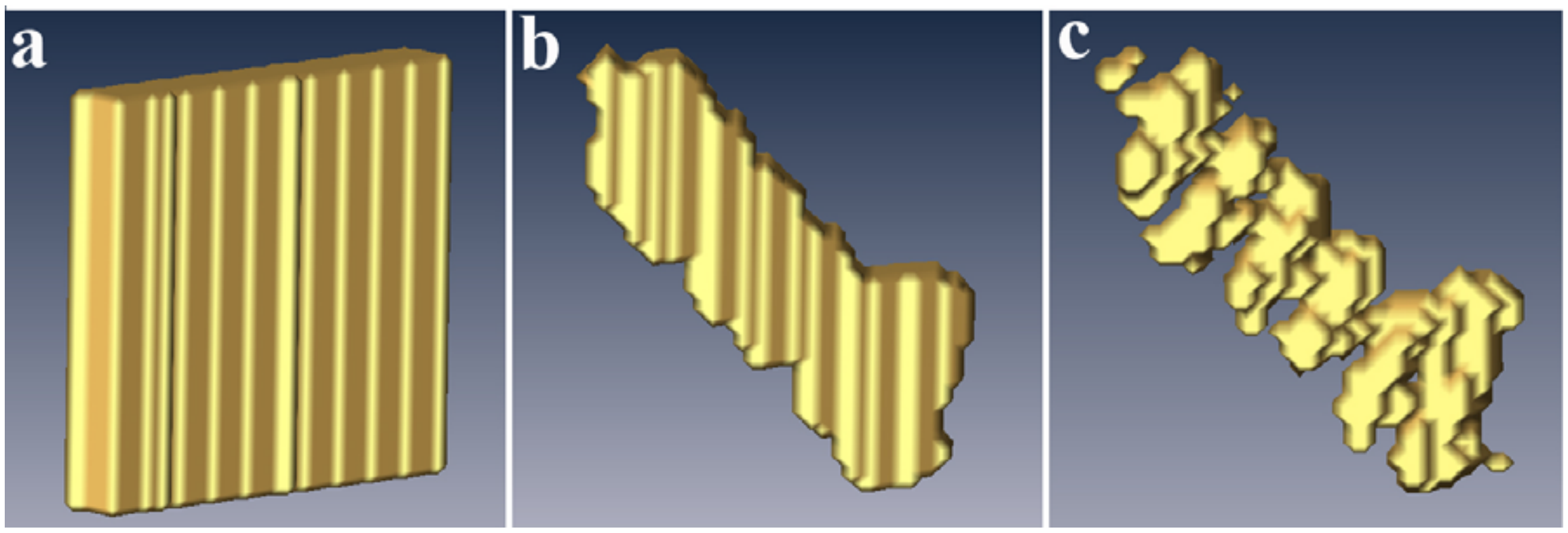

FIG. 3

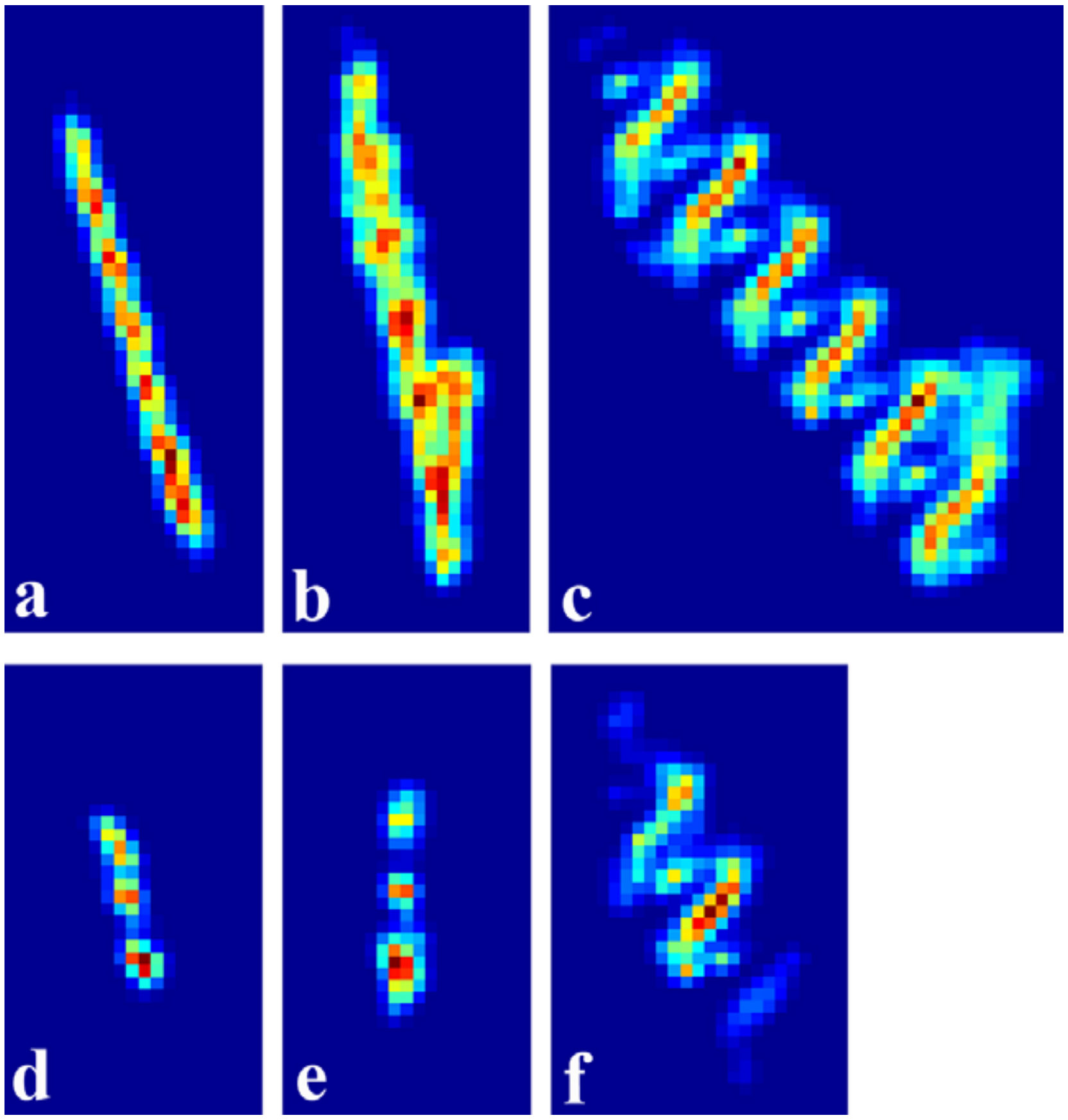

**FIG. 4**

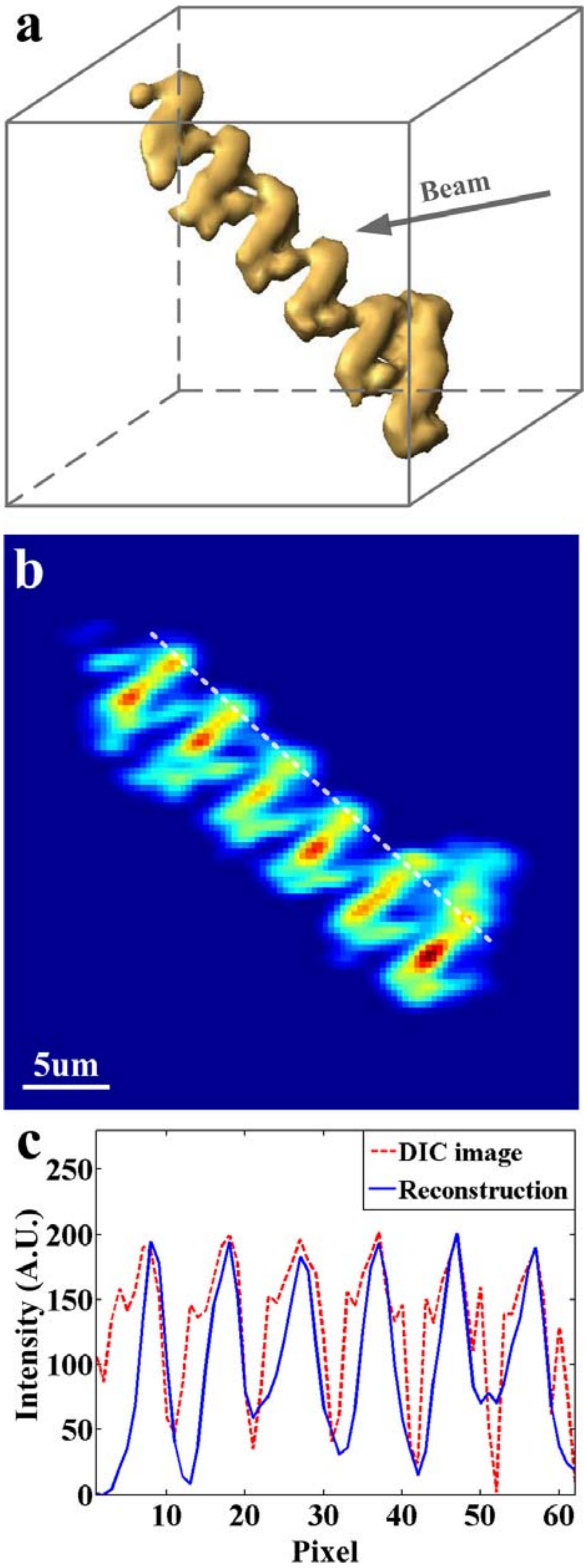

**FIG. 5**